\documentclass[twocolumn,preprintnumbers,amsmath,nofootinbib,amssymb]{revtex4}
\usepackage{amssymb}
\usepackage{lmodern}
\usepackage[utf8x]{inputenc}% For Arxiv Submission Error: Package inputenc Error: Unicode character \xe2\x80\x89 (U+2009) (inputenc) not set up for use with LaTeX
\usepackage[T1]{fontenc}
\usepackage{latexsym}
\usepackage{color}
\usepackage{enumerate}
\usepackage{graphicx}
\usepackage{hyperref}
\usepackage{color}

\def\nn{\nonumber\\}
\newcommand{\f}[2]{\frac{#1}{#2}}
\def\be{\begin{equation}}
\def\ee{\end{equation}}
\def\bea{\begin{eqnarray}}
\def\eea{\end{eqnarray}}
\begin{document}
\title{On Some Applications of the Sagnac Effect}
\author{$^1$A. H. Ziaie\footnote{ah.ziaie@maragheh.ac.ir}, $^1$H. Moradpour\footnote{hn.moradpour@maragheh.ac.ir}, $^2$V. B. Bezerra\footnote{valdir@fisica.ufpb.br}, $^3$A. Jawad\footnote{abduljawad@cuilahore.edu.pk}}
\address{$^1$Research Institute for Astronomy and Astrophysics of Maragha (RIAAM), University of Maragheh, P. O. Box 55136-553, Maragheh,
Iran\\
$^2$ Departamento de F\'isica, CCEN - Universidade Federal da \\
Para\'iba; C.P. 5008, CEP  58.051-970, Jo\~ao Pessoa, PB,
Brasil\\
$^3$ Department of Mathematics, COMSATS University Islamabad,
Lahore-Campus, Lahore, 54000, Pakistan}
\begin{abstract}
Considering exact spacetimes representing rotating black holes and naked singularities, we study the possibility that the Sagnac effect detects $i$) higher dimensions, $ii$) rotation of black holes in higher dimension, and also,
$iii$) distinguishes black holes from naked singularities. The results indicate that the Sagnac time delay gets affected by the presence of extra-dimension or its associated angular momentum. This time delay is also different in the spacetime of a naked singularity compared to that of a black hole. Hence, the Sagnac effect may be used as an experiment for better understanding of spacetimes with higher dimensions or those that admit naked singularities.
\end{abstract}

\maketitle
\section{Introduction}

In order to study the so-called Aether, various experiments such
as the well-known Michelson-Morley experiment, and the Sagnac
experiment have been proposed, and employed
\cite{Sagnac:1899,Sagnac:1913,Fizeau:1851}. Circular relative
motion is the backbone of the Sagnac effect which gets along well
with the ideas of testing modified gravities and detecting a
variety of phenomena predicted in general relativity (GR)
\cite{Tartaglia:1998rh,Rizzi:2003dh,Ruggiero:2010np,Ruggiero:2015gha,Zeylikovich:2011zz,Karimov:2017pxc,EPJC,EPL,Benedetto2020}.
Indeed, this experiment is a gravitational counterpart of the
Aharonov-Bohm effect \cite{Sak}, an achievement which helps us
find a simple way to calculate the Sagnac time delay
\cite{Rizzi:2003dh,Ruggiero:2010np,Sak}.

The problem of the number of spacetime dimensions has attracted much attention
during the past decades and is still a notable challenge in modern physics~\cite{dim} and specially in black hole physics~\cite{Emparan2008}; 
the dimensions number and also the probable rotation of black hole in other dimensions
affect the spacetime geometry. In this regard, the Kerr metric or Kerr geometry describing the geometry of empty spacetime around a rotating uncharged axially-symmetric black hole is a widely used metric in the literature~\cite{ker}. This rich solution has inspired physicists to
seek for different types of rotating black holes like BTZ \cite{BTZ},
and Myers-Perry ones~\cite{MP}. These solutions clearly prove the
implications of dimensions other than four on the properties of black hole,
motivating us to study the Sagnac experiment as a tool to examine physical properties of such solutions. This indeed helps us finding a
way to verify the existence of other dimensions.

Our present understanding of black holes in the universe implies
that these compact objects in the universe are formed as the end
state of a typical gravitational collapse of matter. From a
theoretical perspective, a direct consequence of the use of
GR reveals that, a continual gravitational collapse always leads to spacetime singularities necessarily hidden
by an event horizon, according to cosmic censorship hypothesis
(CCH)~\cite{PenCCH}. Hence, no signal emanated from the
neighborhood of singular region can reach distant observers.
However, neither a conclusive proof or disproof nor a concrete
mathematical formulation of CCH has been provided up until now
leaving this hypothesis as one of the most outstanding unresolved
problems in GR. On the other hand, there are great debates on possible inevitability of the spacetime singularities~\cite{nak8}. In this
regard, recent studies of collapse scenarios show that GR field
equations also admit naked singularities, depending on the initial
conditions and different matter configurations~\cite{Joshib,nak0}. In
this line, the possibility of distinguishing black holes from
naked singularities is an attractive challenge
\cite{nak1,nak2,nak3,nak4,nak5,nak6,nak7,nak8}.

Here, we are initially going to address the possibility that the Sagnac
experiment detects the number of spacetime dimensions as well as the rotation of black holes in other
dimensions. This is the task of the subsequent section. The propose of the
third section is to show that the Sagnac experiment also can distinguish black holes from naked singularities. We set the
units so that $c=G=1$.

\section{Sagnac time delay of rotating black holes}

Our aim in the present section is to calculate the Sagnac time
delay for two well-known black hole solutions. Considering the
results obtained in~\cite{Ruggiero:2010np}, we proceed with the
following steps

\begin{itemize}
    \item Without loss of generality we work within equatorial plane by setting $\theta=\pi/2$. Then, we make coordinate transformation on azimuthal coordinate: $\phi\rightarrow\phi+\omega_0t$, where $\omega_0$ is the rotational velocity of the source of signals and $t$ is the proper time as measured by a clock rotating with constant angular velocity $\omega_0$~\cite{Tartaglia:1998rh,Ruggiero:2010np,Ruggiero:2015gha}.
    \item Define four-vector gravitomagnetic potential as: $A_\mu=-g_{\mu0}/g_{00}$, and
    \item Perform the integration
    \begin{equation}
    \Delta t=2\sqrt{-g_{00}}\int_{0}^{2\pi}A_{\phi}d\phi,\label{btzmetric}
    \end{equation}
    in order to evaluate the Sagnac time delay.
\end{itemize}

\par
\begin{figure}
	\begin{center}
		\includegraphics[scale=0.47]{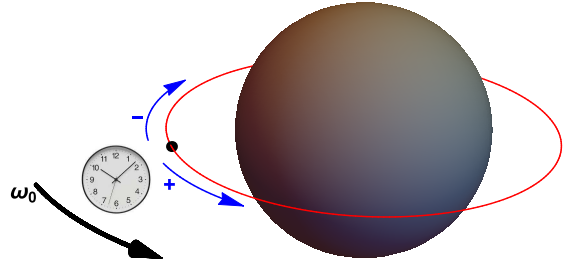}
		\vspace{-4cm}\caption{The time difference as measured by a clock rotating around the black hole with constant angular velocity $\omega_0$ between the emission and absorption of the co-propagating (+) and counter-propagating (-) light beams (red circular orbit). The black point represents a satellite carrying the source/receiver.}\label{fig0}
	\end{center}
\end{figure}

Now, we consider a satellite orbiting in a circular orbit of radius $R = constant$ with constant angular velocity $\omega_0$ around the gravitating body~\cite{Ruggiero:2010np,Tarta2020}. This satellite acts as a source/receiver of light beams propagating along circular orbits of the same radius around a source of gravitational field which here is taken as a Kerr black hole, see Fig.~(\ref{fig0}). To calculate the Sagnac time delay for this black hole, we begin with its line element given by
\bea\label{Kmetric}
ds^2&=&-\left[1-\f{2Mr}{r^2+a^2\cos^2\theta}\right]dt^2+\f{r^2+a^2\cos^2\theta}{a^2-2Mr+r^2}dr^2\nn&+&\left(r^2+a^2\cos^2\theta\right)d\theta^2-\f{4Mra\sin^2\theta}{r^2+a^2\cos^2\theta}dt
d\phi\nn&+&\left[r^2+a^2+\f{2Mra^2\sin^2\theta}{r^2+a^2\cos^2\theta}\right]d\phi^2,
\eea where $a=J/M$ is the angular momentum per black hole mass. A
straightforward calculation then reveals that the Sagnac time
delay for a Kerr black hole is obtained as \be\label{timeKErr}
\Delta t_{\rm K}=\f{4\pi\left[\omega_0\left(Ra^2+2Ma^2+R^3\right)-2aM\right]}{R\left[1-\f{2M}{R}+\f{4aM\omega_0}{R}
-\omega_0^2\left(\f{2Ma^2}{R}+a^2+R^2\right)\right]^{\f{1}{2}}}.
\ee This time difference, as measured by a sensitive clock which is carried by the satellite corresponds to the round
trip time for a co-propagating light beam traveling along a
circular orbit. For a counter-propagating beam we must set
$a\rightarrow-a$. It can be also verified that in the limit
$M\rightarrow0$, $a\rightarrow0$, the Sagnac time delay reduces to
its counterpart in 4-dimensional Minkowski spacetime. We note that
the condition on $\Delta t_{\rm K}$ to be a real and positive
number requires that the numerator of expression (\ref{timeKErr})
be positive and its denominator assumes real numbers. We further
note that the denominator of expression (\ref{timeKErr}) has two
roots given by \bea\label{timeKErrroots}
a_-&=&\f{2m\omega_0-\sqrt{R^2\omega_0^2-2mR^3\omega_0^4-R^4\omega_0^4}}{2m\omega_0^2+R\omega_0^2},\nn
a_+&=&\f{2m\omega_0+\sqrt{R^2\omega_0^2-2mR^3\omega_0^4-R^4\omega_0^4}}{2m\omega_0^2+R\omega_0^2}.
\eea
\begin{figure}
    \begin{center}
        \includegraphics[scale=0.47]{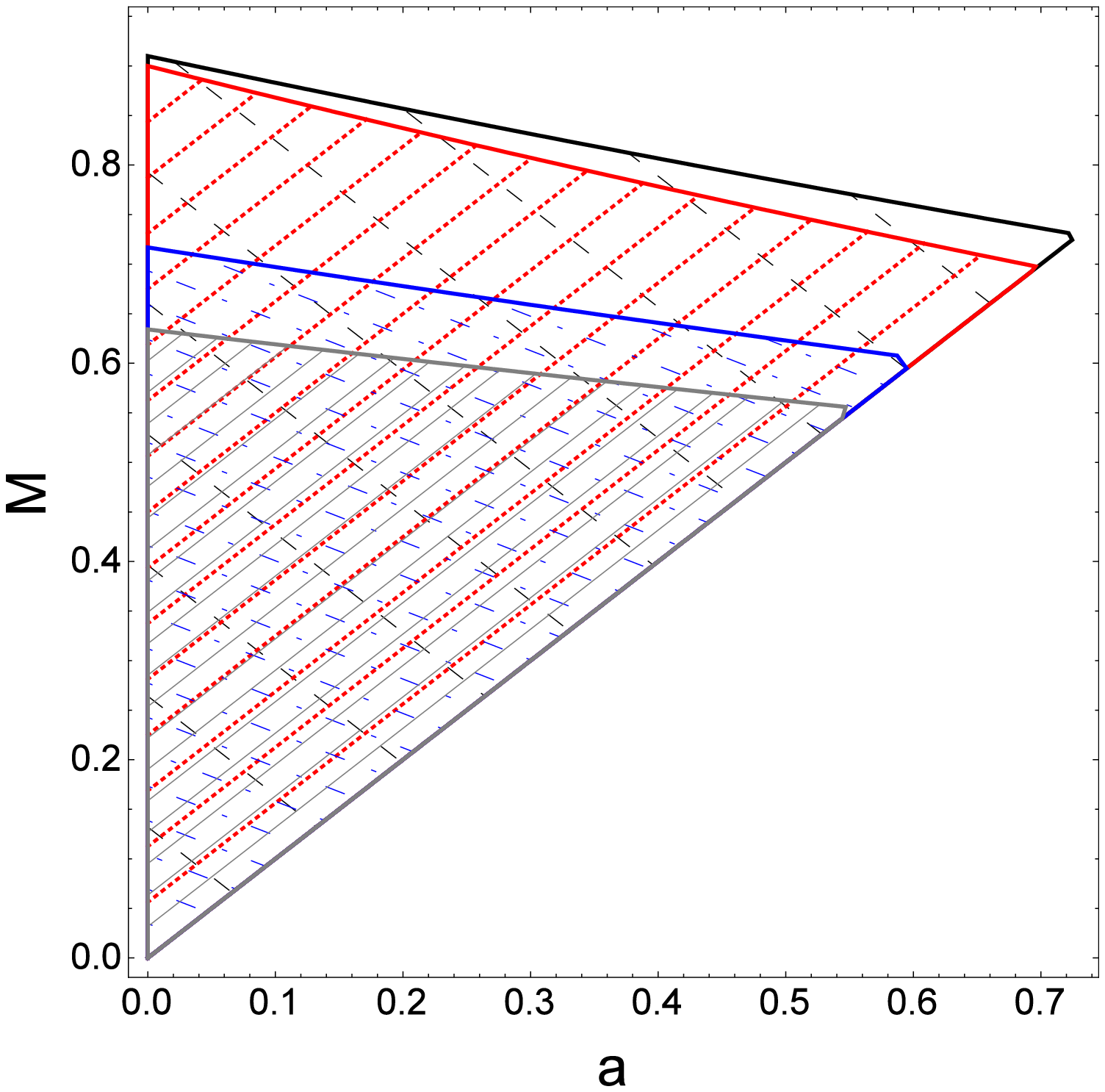}
        \includegraphics[scale=0.47]{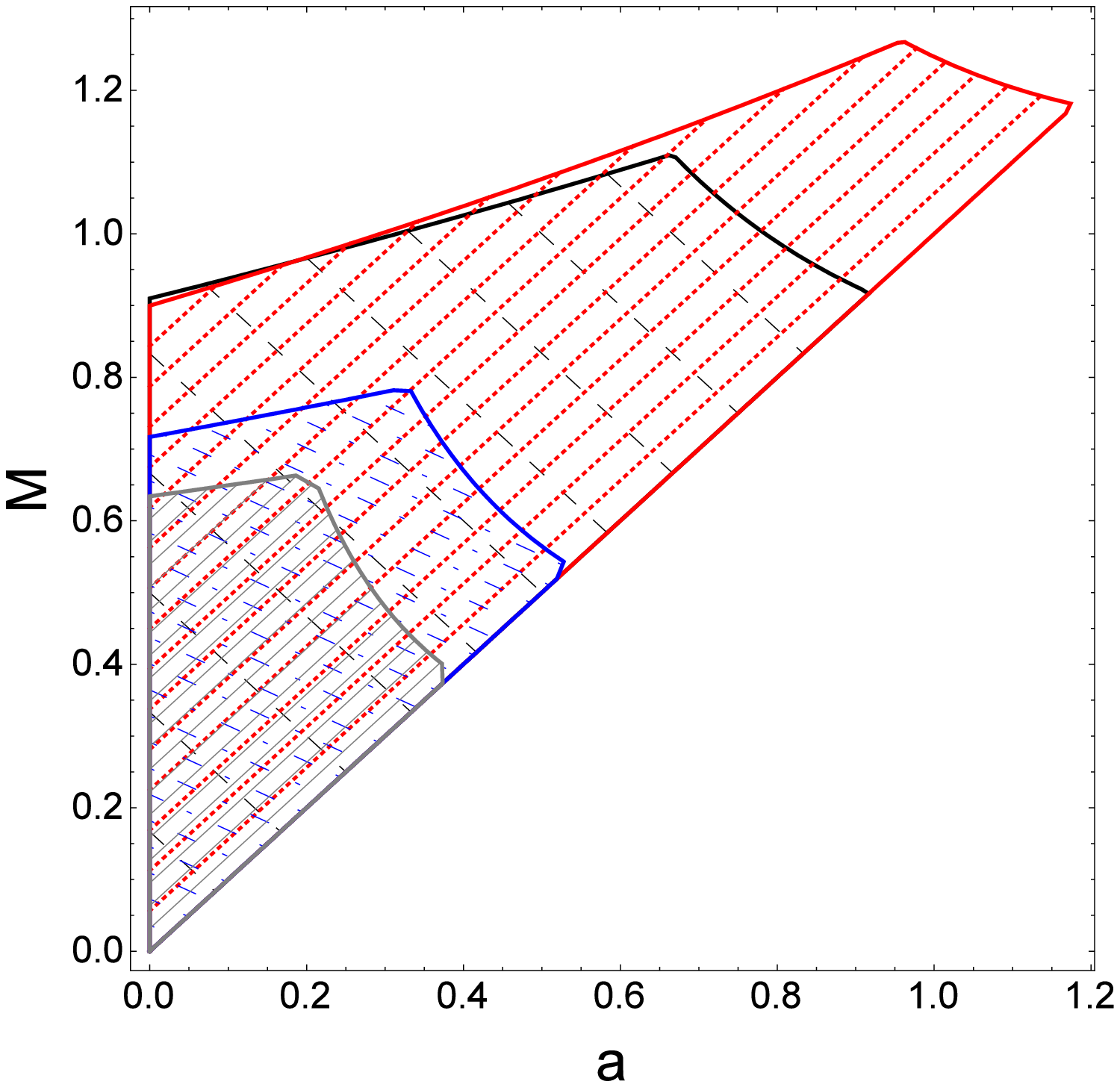}
        \caption{The allowed regions of mass and angular momentum for $\omega_0=0.15$ and $R=2.0$ (black region), $\omega_0=0.18$ and $R=2.1$ (red region), $\omega_0=0.14$ and $R=1.5$ (blue region) and $\omega_0=0.12$ and $R=1.3$ (gray region). The upper (lower) panel stands for counter (co)-propagating beams.}\label{fig1}
    \end{center}
\end{figure}
Hence for $\Delta t_{\rm K}$ to be finite we require that the black hole angular momentum satisfies the condition $a_-<a<a_+$. In Fig. (\ref{fig1}) we have sketched the allowed regions for mass and angular momentum satisfying the above conditions. The upper and lower panels present the allowed regions for counter and co-propagating beams, respectively. Figure (\ref{fig2}) presents the behavior of $\Delta t_{\rm K}$ against black hole mass. We observe that for co-propagating beams, as the black hole angular momentum increases the Sagnac time delay decreases and such a decrement depends on black hole mass. In fact, the rotation of black hole assists shortening of the light path as a result of frame dragging effect. For counter-propagating beams, the faster the black hole rotates, the longer its takes for the beam to complete its orbit.
In the limit $a\rightarrow0$ we still have non-vanishing time delay which is due to nonzero rotational velocity of the source, however, the more the black hole weight, the greater the Sagnac time delay.
\begin{figure}
    \begin{center}
        \includegraphics[scale=0.5]{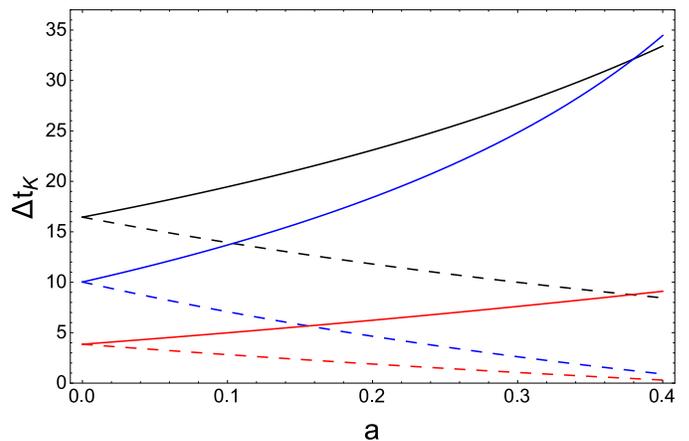}
        \caption{Behavior of Sagnac time delay versus angular momentum for $\omega_0$ and $R$ chosen from regions of Fig. (\ref{fig1}) and different values of Kerr black hole mass. The family of solid curves  represent the counter-propagating and those of dashed ones represent the co-propagating beams. For the model parameters we have set $\omega_0=0.15$, $R=2.0$ and $M=0.7$ (black curves), $\omega_0=0.14$, $R=1.5$ and $M=0.6$ (blue curves) and $\omega_0=0.12$, $R=1.3$ and $M=0.35$ (red curves).}\label{fig2}
    \end{center}
\end{figure}
\par
Next, we proceed to calculate the Sagnac time delay for a
5-dimensional black hole known as Myers-Perry black hole whose
line element is given by \bea\label{MPmetric}
ds^2&=&-dt^2+\f{M\left(dt+a\sin^2\theta d\phi+b\cos^2\theta
d\psi\right)^2}{r^2+a^2\cos^2\theta+b^2\sin^2\theta}\nn&+&\f{r^2\left(r^2+a^2\cos^2\theta+b^2\sin^2\theta\right)}{(r^2+a^2)(r^2+b^2)-M
r^2}dr^2\nn&+&(r^2+a^2)\sin^2\theta d\phi^2+(r^2+b^2)\cos^2\theta
d\psi^2\nn&+&\left(r^2+a^2\cos^2\theta+b^2\sin^2\theta\right)d\theta^2,
\eea where $b$ is the angular momentum of fifth-dimension. Similarly
to the above procedure we get the Sagnac time delay as \bea
\Delta t_{\rm MP}&=&\f{4\pi}{\left(R^2+b^2\right)^{\f{1}{2}}}\times\nn\!\!\!\!\!\!&&\f{\left[\omega_0\left(2M
a^2+a^2b^2+R^2b^2+R^2a^2+R^4\right)-2aM\right]}{\left[R^2+b^2-2M+{\mathcal
G}\right]^{\f{1}{2}}},\nn \eea where \be {\mathcal
G}=\omega_0\left[4aM-\omega_0(2M
a^2+a^2b^2+R^2b^2+R^2a^2+R^4)\right]. \ee
\begin{figure}
    \begin{center}
        \includegraphics[scale=0.5]{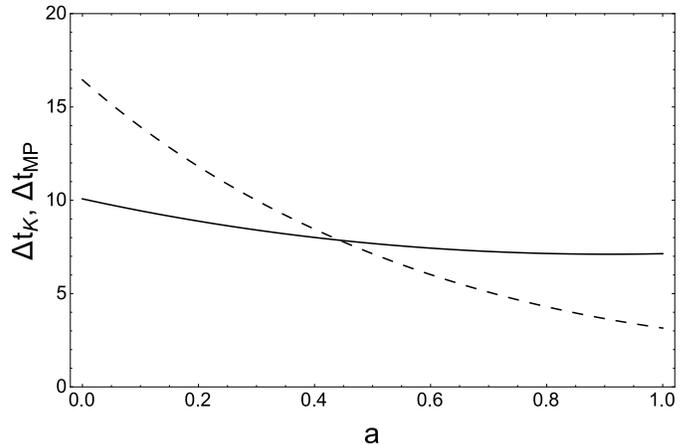}
        \caption{Behavior of Sagnac time delay for Kerr (dashed curve) and Myers-Perry (solid curve) black holes versus angular momentum of fourth-dimension. For the model parameters we have set $b=0$, $\omega_0=0.15$, $R=2.0$ and $M=0.7$. The critical value of angular momentum is found as $a_{\rm cr}=0.4428$.}\label{fig3}
    \end{center}
\end{figure}
We note that in the limit $a=b\rightarrow0$, $M\rightarrow0$, the  above expression reduces to Sagnac time delay in 5-dimensional Minkowski spacetime. At first glance we observe that even if we set $b=0$, the Sagnac time delay of a Kerr black hole is different to that of Myers-Perry whose angular momentum in fifth dimension is zero. This means that even if there is no rotation in the fifth dimension, the round trip time for a 4-dimensional black hole, as given by Eq. (\ref{timeKErr}) is different from a 5-dimensional black hole, due to the presence of extra dimension. This has been shown in Fig. (\ref{fig3}) where we have plotted the two time delays against the angular momentum of the fourth-dimension. As we observe, both time delays are decreasing functions of angular momentum, however, for the case of Kerr black hole, the Sagnac time delay decreases at a higher rate, see dashed curve. There is a critical value for the angular momentum of fourth-dimension ($a_{\rm cr}$) at which $\Delta t_{\rm K}=\Delta t_{\rm MP}$. For $a<a_{\rm cr}$ we have $\Delta t_{\rm K}>\Delta t_{\rm MP}$ and vice versa. Also in the limit $a\rightarrow0$ both time delays are non zero due to the angular velocity of the source/receiver but $\Delta t_{\rm MP}(a=0)<\Delta t_{\rm K}(a=0)$. Figure (\ref{fig4}) shows a comparison of Sagnac time delay for Kerr and Myers-Perry black holes for different values of angular momentum of fifth-dimension. Interestingly, we observe that increasing the $b$ parameter shifts $\Delta t_{\rm MP}$ from decreasing to increasing function of angular momentum of the fourth-dimension. This means that the faster the black hole rotates in fifth dimension the more chance we may have to distinguish it, through the Sagnac effect, from a purely 4-dimensional Kerr black hole.
\begin{figure}
    \begin{center}
        \includegraphics[scale=0.5]{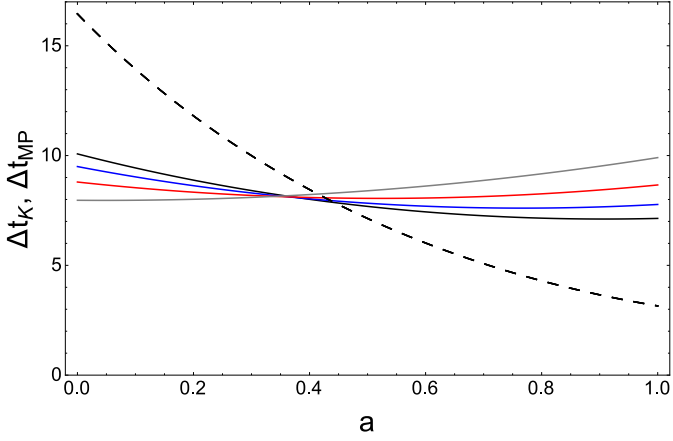}
        \caption{Behavior of Sagnac time delay for Kerr (dashed curve) and Myers-Perry (solid curves) black holes versus angular momentum of fourth-dimension. For the model parameters we have set $b=0$ (black curve), $b=1$ (blue curve), $b=2$ (red curve), $b=10$ (gray curve), $\omega_0=0.15$, $R=2.0$ and $M=0.7$.}\label{fig4}
    \end{center}
\end{figure}

\section{Discriminating between black hole and naked singularities}

Our aim in the present section is to employ the Sagnac effect as a
tool to distinguish between a black hole and naked singularity.
Let us begin with the line element for a rotating naked
singularity which in Boyer-Lindquist coordinates is given
by~\cite{prd2008} \bea\label{nsmetric}
ds^2&=&-\f{\Delta-a^2\sin^2\theta}{\Sigma}dt^2+\f{\Sigma}{\Delta}dr^2+\Sigma
d\theta^2\nn&+&\sin^2\theta\left[\Sigma+
\left(2-\f{\Delta-a^2\sin^2\theta}{\Sigma}\right)a^2\sin^2\theta\right]d\phi^2\nn&-&2\left(1-\f{\Delta-a^2\sin^2\theta}{\Sigma}\right)a\sin^2\theta
dtd\phi, \eea where \bea\label{defs}
&&\Sigma=h^{1-\mu}\rho,~~\Delta=r^2+a^2-\f{2Mr}{\mu},~~\rho=r^2+a^2\cos^2\theta,\nn
&&\mu=\f{M}{\sqrt{M^2+q^2}},~~h=1-\f{2Mr}{\mu\rho}, \eea and $M$
and $q$ are the Arnowitt-Deser-Misner (ADM) mass and scalar charge
and are constant real parameters. The above metric is the rotating
version of Janis-Newman-Winicour solution~\cite{JNW} and
represents the spacetime of a rotating globally naked singularity
(for $0\leq\mu<1$)~\cite{ROTGNS} in the presence of a massless
scalar field whose radial dependence is given by \bea\label{sf}
\Phi(r)=\f{\sqrt{1-\mu^2}}{4}\ln\left(1-\f{2Mr}{\mu\rho}\right).
\eea It can be easily seen that for $\mu=1$ (or equivalently
$q=0$) the Kerr metric is recovered.
\par
\begin{figure}
    \begin{center}
        \includegraphics[scale=0.5]{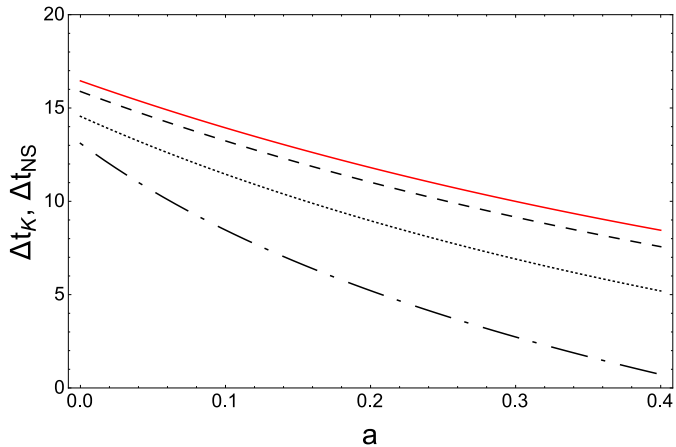}
        \caption{Behavior of Sagnac time delay for Kerr black hole (red curve) and a rotating naked singularity (black curves) versus angular momentum. For the model parameters we have set $\omega_0=0.15$, $R=2.0$, $M=0.7$ and $\mu=0.95$ (dashed curve), $\mu=0.85$ (dotted curve), $\mu=0.75$ (long-dashed curve).}\label{fig5}
    \end{center}
\end{figure}

Now, assume that the source/receiver orbits with angular velocity
$\omega_0$ at constant radius $r=R$ in the plane $\sin\theta=k$.
Simple calculations then give \be\label{ds2NS}
ds^2=\f{k^2}{B}\left(Dd\phi-adt\right)^2-\f{A}{B}\left(dt-ak^2d\phi\right)^2,
\ee where \be\label{DEFS}
A=\Delta\Big|_{r=R},~~~B=\Sigma\Big|_{r=R}^{\sin\theta=k},~~~D=\left(\Sigma+a^2k^2\right)\Big|_{r=R}^{\theta=0}.
\ee If we denote the angular velocity of light beam as $\Omega$,
then $\phi=\Omega t$. The angular velocity of light rays is then
obtained as \be\label{ANGVL}
\Omega_\pm=\f{a\pm\f{\sqrt{A}}{k}}{D\pm ak\sqrt{A}}, \ee where use
has been made of the condition $ds^2=0$ for null geodesics. We
note that the above result is valid within the planes for which
$D\pm ak\sqrt{A}\neq0$. During the orbital rotation of the light
beams, the source/receiver rotates at angular distance $\phi_0$
and consequently, when the two light beams reach the
source/receiver they experience phase difference, depending on the
relative motion of the light beams with respect to the
source/receiver. Indeed, the light beam which is
co(counter)-rotating with the source/receiver, will travel at
angular distance $\phi_0+2\pi$ ($\phi_0-2\pi$). Thus, for the
distances that the light beams travel we have
\be\label{disdegrees}
\f{\Omega_\pm}{\omega_0}\phi_{0\pm}=\phi_{0\pm}\pm2\pi, \ee whence
we get \be\label{disdegreesf}
\phi_{0\pm}=\pm\f{2\pi\omega_0}{\Omega_\pm-\omega_0}. \ee We note
that $\phi_{0\pm}$ is the angular distance travelled with the
light beam with angular velocity $\Omega_{\pm}$. From
Eq.(\ref{ds2NS}) we get the following time difference for the
source/receiver as \be\label{dtsr}
dt=\f{d\phi}{\omega_0\sqrt{B}}\left[k^2\left(D\omega_0-a\right)^2-A\left(1-ak^2\omega_0\right)^2\right]^{\f{1}{2}}.
\ee Finally, integrating from $\phi_{0-}$ to $\phi_{0+}$ along
with utilizing expression (\ref{disdegrees}) we can find the
Sagnac time difference between the two light signals. For $k=1$,
the Sagnac time delay reads \be\label{sagns} \Delta t_{\rm
NS}=\f{4\pi\left[\omega_0\left(D^2-a^2A\right)-a(D-A)\right]}{Rh^{\f{1-\mu}{2}}\sqrt{A(1-a\omega_0)^2-(D\omega_0-a)^2}}.
\ee It can be checked that for $\mu=1$ the above relation reduces
to the Sagnac time delay for a Kerr spacetime. In Fig.
(\ref{fig5}) we have compared the Sagnac time delay for a Kerr
black hole and its naked singularity counterpart. As we observe,
both time delays are decreasing functions of angular momentum,
however, the time delay for naked singularity decreases more
rapidly for smaller values of $\mu$ parameter. Thus, for a
specific angular momentum, the lesser the value of $\mu$
parameter, the more chance we have to distinguish a naked
singularity from a black hole with same angular momentum.

\section{Summary and Concluding remarks}
Motivated by the idea that if spacetime admits extra-dimensions, then its implications should appear in gravitational
measurements around black holes, in the present work, we employed the Kerr and the Myers-Perry metrics
and studied the capability of Sagnac effect in recognizing possible existence of the fifth
dimension as well as the rotation of black holes in this dimension. To this aim, we studied the interference process of light beams in a rotating frame in curved spacetime. The corresponding set up is based on an interferometer carried by a satellite that uniformly rotates in a circular orbit around the black hole. This satellite also carries a clock that measures the difference between propagation time of each of the co-propagating and counter-propagating beams~\cite{Ruggiero:2010np,Tarta2020}.

\par 
We then observed that even if black hole has no rotation in fifth dimension, the Sagnac time delay of a Kerr black hole will be different from that of Myers-Perry due to the mere presence of the fifth dimension. We also found that the change in angular momentum of the fifth dimension affects the Sagnac time delay so that it could be distinguished from a four-dimensional Kerr black hole. Although, our focus was on four and five dimensional cases, the present research can legally be extended to higher dimensions. 
\par
Moreover, using the Janis-Newman-Winicour metric which represents the spacetime of a rotating naked singularity, it is shown that the Sagnac effect may also be used to distinguish naked rotating singularities
from a Kerr black hole. The outcomes of the present article confirm the power of Sagnac effect in studying possible existence of extra-dimensions and naked singularities in the Universe. 
\section*{Acknowledgments}
V.B.B. is partially supported by the National Council for Scientific and Technological Development – CNPq (Brazil), through the Research Project No. 307211/2020-7. This paper is published as part of a research project supported by the University of Maragheh Research Affairs Office.
\end{document}